\documentclass{elsart}

 \usepackage{epsfig}

\usepackage{amssymb}

\begin{document}

\begin{frontmatter}



 \corauth[cor1]{Corresponding author:
V. Dobrosavljevi\'c, fax. 850-644-5393}

\title{Glassy Behavior of Electrons as a Precursor to the
 Localization Transition}


\author[1]{V. Dobrosavljevi\'c}
\ead{vlad@magnet.fsu.edu}
\author[1]{A. A. Pastor}
\address[1]{Department of Physics and 
National High Magnetic Field Laboratory\\ Florida State
University, 1800E Paul Dirac Dr., Tallahassee, FL 32310, USA}
\begin{abstract}
A theoretical model is presented, describing the glassy freezing of 
electrons in the vicinity of disorder driven metal-insulator transitions. 
Our results indicate that the onset of glassy dynamics should emerge 
{\em before} the localization transition is reached, thus predicting
the existence of an intermediate metallic glass phase between the 
normal metal and the insulator. 
\end{abstract}

\end{frontmatter}

\section{Introduction}

Disorder-driven metal-insulator transitions (MITs) in strongly
correlated electronic systems continue to demonstrate fascinating
behaviors that remain to be elucidated.  The competition of
electron-electron interactions and disorder is most pronounced in the
insulator, where one generally expects the existence of a large number
of low energy configurations of the electronic density, leading to
glassy behavior, metastability, and slow relaxation.  Far from the
transition, the kinetic energy of the electrons is negligible and one
can utilize a classical picture, as pioneered in early work by Efros
and Shklovskii \cite{efros}. The precise role of glassy dynamics is
more difficult to assess closer to the MIT, where quantum fluctuations
due to electronic mobility become important. In recent work
\cite{pastor}, we have developed a formalism capable of describing
such glassy behavior of electrons both in the classical limit and in
presence of quantum fluctuations. Here we extend this
approach to incorporate Anderson localization effects \cite{anderson},
which prove to have a dramatic effect on the stability of the glass
phase.

The following physical picture emerges from our study: Anderson
localization leads to the formation of bound electronic states, thus
suppressing quantum fluctuations introduced by mobile
electrons. However, we find that glassy behavior is {\em not}
eliminated as soon as the electrons become delocalized. Very close to
the MIT the electronic mobility remains very small, thus limiting the
{\em size} of quantum fluctuations needed to overcome barriers between
metastable states. An intermediate metallic glass phase is predicted,
separating the insulator and the normal metal.

\section{Model}

We consider the simplest microscopic model capable of describing the
interplay of glassy behavior and Anderson localization effects. It is
given by a one-band tight-binding model of spinless electrons at half
filling, with nearest-neighbor hopping $t_{ij}$, interacting by infinite
range inter-site interactions $V_{ij}$, in presence of random site
energies $\varepsilon_i$. For technical simplicity, these interactions
are chosen also to be random \cite{random}, with zero mean and
variance $<V_{ij}^2 > =$ $V^2 /N$, where $N$ is the number of sites in
the systems. In the following, we briefly outline the formalism that
we use; the details will be presented elsewhere \cite{prllocgl}.  We
first formally average over random interactions using the replica
method \cite{pastor}, which generates a four-density interaction
term. This term is then decoupled by introducing Hubbard-Stratonovich
fields, and the resulting action takes the form
$S_{eff} = S_{el} + S_{int}$, where

\begin{equation}
S_{el\;\;\;} = \int_o^{\beta}d\tau \sum_{a}\sum_{<ij>}
c_i^{\; \dagger a} (\tau )[ (\partial_{\tau} +\varepsilon_i )
\delta_{ij}
-t_{ij} ]c_i^{\; a} (\tau ),\end{equation}\vspace{-24pt}
\begin{eqnarray}
S_{int\;} = V\sum_i \int_o^{\beta }d\tau \int_o^{\beta }d\tau '
& & \left[  \sum_{a} \tilde{Q}^{aa}_i 
(\tau , \tau ') \delta n_i^{a}(\tau )\delta n_i^{a} (\tau ')\right.\nonumber\\
& & +\left.
\sum_{a < b} Q^{ab}_i 
(\tau , \tau ') \delta n_i^{a}(\tau )\delta n_i^{b} (\tau ')\right].
\end{eqnarray}
We have used standard functional integration over replicated Grassmann
fields, where $a=1,...,n \; (n\rightarrow 0)$ are the replica
indices \cite{pastor}.  Here, the operators $\delta n_{i}^{a}(\tau )=(
c^{\dagger a}_{i} (\tau )c_{i}^{a} (\tau ) - 1/2)$ represent the {\em
density fluctuations} from half filling.

In the classical limit ($t_{ij} =0$), these equations can be solved
exactly due to the infinite range of the interactions $V_{ij}$,
but once finite-range hopping is introduced, the resulting 
electronic degrees of freedom generate additional effective 
nonlocal interactions between density modes, and we have to
resort to approximations. A mean-field treatment of the interactions is then 
provided by evaluating the $Q$-fields in the saddle-point approximation, 
which also makes it possible to identify the replica symmetry breaking 
instability used to establish the onset of glassy ordering \cite{pastor}.

The resulting instability criterion takes the form
\begin{equation}
1=  V^2 < \frac{1}{N}\sum_{ij} (\chi_{ij} )^2 >_{dis}.
\end{equation}
Here, the non-local compressibilities $\chi_{ij}$ are defined
by
\begin{equation}
\chi_{ij} = \frac{1}{\beta}\int_o^{\beta }d\tau \int_o^{\beta }d\tau '
\left[ < \delta n_i^{a}(\tau )\delta n_j^{a} (\tau ')>_o -
<\delta n_i^{a}(\tau )>_o <\delta n_j^{a} (\tau ')>_o \right],
\end{equation}
where the quantum averages $<\cdots >_o$ are taken with respect to the 
full replica-symmetric action $S_{eff}$ evaluated at the saddle-point,
and $<\cdots >_{dis}$ indicates the average with respect to disorder.

Further simplification is obtained \cite{pastor} by concentrating on
the limit of large disorder, where to leading order one can evaluate
the compressibilities by setting $V=0$ in $S_{eff}$. The calculation
then reduces to calculating the compressibilities with respect to a
system of noninteracting electrons in presence of disorder.
Physically, this calculation amounts to focusing on  electrons in the
vicinity of the Anderson transition, and examining the
leading perturbation introduced by turning on inter-site interactions
$V_{ij}$. The crucial insight is obtained by observing that in
contrast to the average compressibilities $<\chi_{ij}>_{dis}$, the
quantities $<(\chi_{ij} )^2>_{dis}$ actually {\em diverge } at $T=0$
at the Anderson transition. We immediately conclude that
Anderson localization introduces a singular perturbation to the
stability of the glass phase, as anticipated in Ref. \cite{pastor}. 
In addition, we note that the sum in Eq. (3) runs over
positive definite terms, so that a {\em lower bound} to the size of
the glass phase is obtained by retaining only the local ($i=j$)
term. We have examined this quantity using several different
approaches to Anderson localization, but the simplest closed form
expression can be obtained by using the recently developed {\em
typical medium theory} \cite{tmt}.

We have used   a simple semi-circular model density of states
\cite{pastor} in presence of a uniform distribution of random 
site energies of width $W$, and have done calculations as a function
of the disorder strength $W$ and temperature $T$. We find that
$<(\chi_{ii} )^2>_{dis} \sim (W_c -W)^{-1}$ indeed blows up 
at $T=0$ precisely
at the Anderson transition, which for this model occurs at $W_c
\approx 1.36 B$, where $B$ is the electronic bandwidth.  As a result,
the instability criterion for the emergence of the glass phase is
satisfied {\em before} the Anderson transition is reached, for any
finite value of the interaction $V$.  To illustrate our findings, we
present the resulting phase diagram for $W/V =2$, as a function of $W$
and $T$ in Fig. 1. Note that the ``kink'' in the glass transition
boundary at $W=W_c$ (full line) reflects the crudeness of our treatment
of localization, which ignores the fact that inelastic scattering
effects will tend to delocalize the electrons at any finite
temperature. These inelastic effects can be accounted for by keeping
the interaction terms in $S_{eff}$, but these corrections will be
investigated in more detail elsewhere. To illustrate the qualitative
effects of such inelastic scattering, we phenomenologically introduce
a temperature-dependent scattering rate of the form $\eta = A T$. The
resulting modification of the phase boundary for the choice $A=0.1$ is
presented by a dotted line, leading to the rounding of the kink, as
expected.

\section{Discussion}

Our results strongly suggest that the onset of glassy dynamics should
be considered as a {\em precursor} to disorder-driven metal-insulator
transitions.  Physically, as the system approaches localization the
electrons are barely mobile, allowing for even moderate interactions
to induce meta-stability and glassy freezing. We do not expect this
glassy freezing to have a direct effect on the {\em average}
conductivity, but it should be more readily identified by examining
the slowing down of the electron dynamics. In particular, as the
glassy phase is entered, we expect large enhancements for the low
frequency components of the conductivity noise spectra, reflecting
slow transitions between emerging metastable states. An interesting
question relates to the the {\em size} of the proposed metallic glass
phase as a function of disorder. This intermediate phase is expected
to be extremely small for very strong disorder, since in this limit
the effect of the interactions leading to glassy behavior should be
very modest.  But what should be expected for very weak disorder but
strong interactions?  In the limit where $r_s = V/E_F >> 1$,
insulating behavior can emerge due to Wigner crystallization, even in
absence of disorder. As the Wigner crystal is approached, we expect
the local compressibility to be strongly reduced, reflecting strong
short range order induced by electronic correlations. As a result, the
possibility for many electronic configurations is again impaired,
suppressing glassy ordering. We thus expect that the best chance to
observe the metallic glass phase is for moderately disordered systems
with $W/V \sim 1$. Remarkably, very recent experiment on a
two-dimensional electron gas in silicon seems to support this picture.
In this material, clear evidence for an intermediate metallic glass
phase has been discovered for low mobility samples
\cite{snezana}, but {\em not} in high-mobility devices. In
latter systems, the emergence of glassy dynamics seems to {\em
coincide} with the MIT \cite{jan}, as one would expect for systems
closer to a (disordered) Wigner crystal.

 This work was supported by the NSF grant DMR-9974311 and
the National High Magnetic Field Laboratory. We thank
S. Bogdanovich, J.\ Jaroszy\'nski, and D. Popovi\'c for useful
and stimulating discussions.

\newpage

\section{Figure captions}

Figure 1

Phase diagram for spinless interacting electrons 
in presence of disorder, as a function of temperature $T$ and the
Fermi energy $E_F$, expressed in units of the disorder strength $W$.
Results are presented for moderate interaction strength $W/V =2$. The
glass transition boundary is shown, in absence of inelastic scattering
(full line) and in presence of an inelastic scattering rate $\eta=0.1
T$ (dotted line). Note that the glassy phase emerges {\em before} the
metal-insulator transition (MIT) is reached (heavy dot). The (glassy)
Anderson insulator is shown with a heavy full line (at $T=0$).
\vfill
\pagebreak

\begin{figure}[h]
\centerline{\epsfig{file=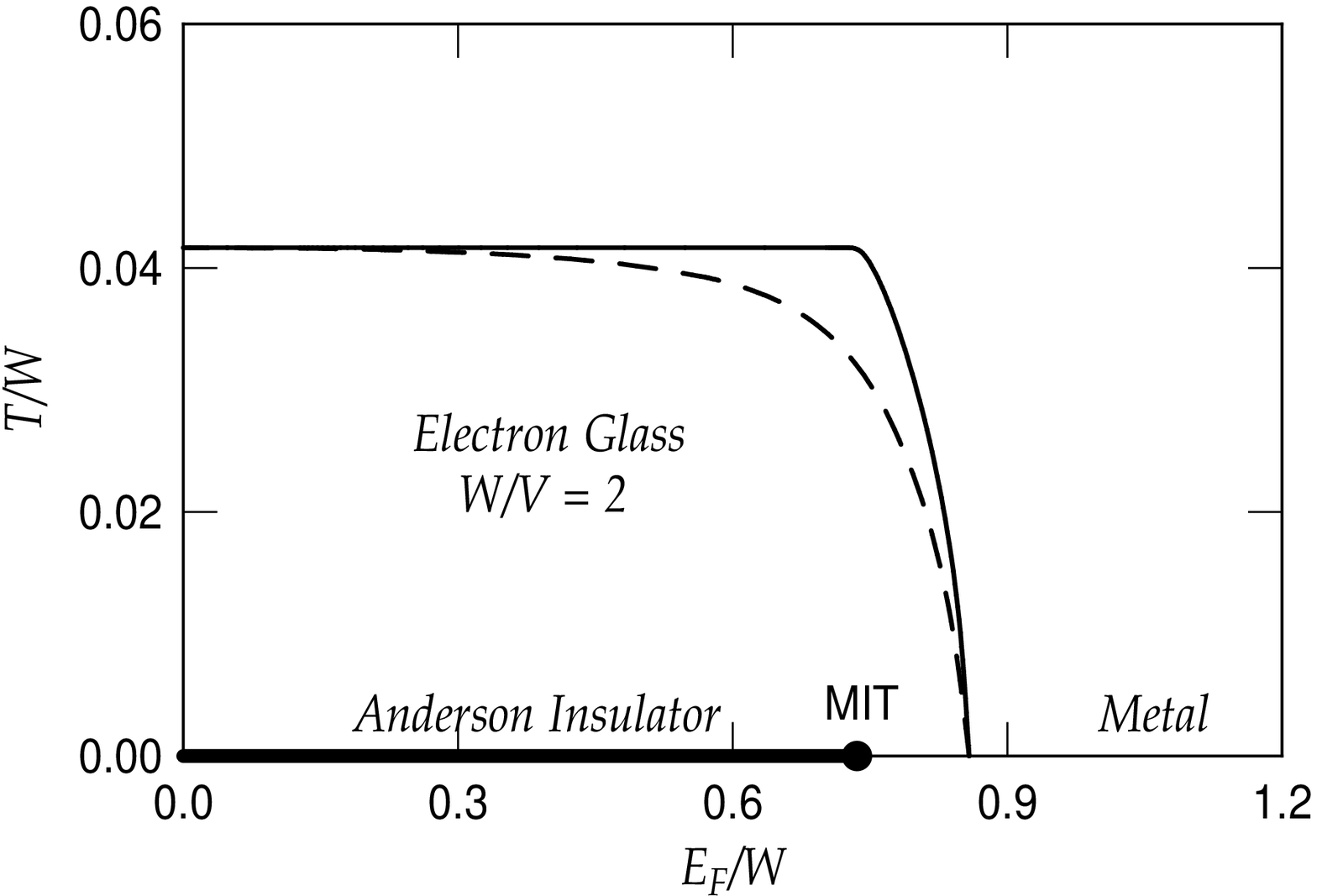,width=6in}}
\end{figure}


\begin{thebibliography}{99}\parsep 0pt\parskip -0pt \itemsep 2pt
\bibitem{mott}  N. F. Mott, 1990
{\em Metal-Insulator Transitions} (London: Taylor and Francis, 1990).

\bibitem{efros} A. L. Efros and B. I.  Shklovskii, J. Phys. C{\bf 8} 
L49(1975).

\bibitem{pastor}
A. A. Pastor and V. Dobrosavljevi\'c,
Phys. Rev. Lett. {\bf 83}, 4642 (1999). 

\bibitem{anderson}
P. W. Anderson, Phys. Rev. {\bf 109}, 1498 (1958)

\bibitem{random} In recent work \cite{pastor}, we have shown that 
in presence of random site energies, such random interactions are generated
by renormalization even if one starts with purely repulsive 
interactions in the bare Hamiltonian. Similar conclusions have been
obtained numerically in: E. R. Grannan and C. C. Yu, Phys. Rev. Lett.
{\bf 71}, 3335 (1993).

\bibitem{prllocgl} V. Dobrosavljevi\'c and A. A. Pastor,
in preparation.

\bibitem{tmt} V. Dobrosavljevi\'c and A. A. Pastor,
preprint, cond-mat/0106282.

\bibitem{snezana} S. Bogdanovich and D. Popovi\'c, proceedings of EP2DS14,
Physica E (2002).
\bibitem{jan} J.\ Jaroszy\'nski, D. Popovi\'c, and T. M. Klapwijk,
 proceedings of EP2DS14, Physica E (2002).


\end{thebibliography}
\end{document}